\documentclass[11pt,a4paper]{article}

\usepackage{amsmath}
\usepackage{amsfonts}
\usepackage{amssymb}
\usepackage{caption}
\usepackage[font={color=LIGHTBLUE},figurename=Figure.,labelfont={bf}]{caption}
\usepackage{color}
\usepackage[english]{babel}
\usepackage{enumitem}
\usepackage{fancybox}
\usepackage{fancyhdr}
\usepackage{fchicago} 
\usepackage[Conny]{fncychap}
\usepackage[T1]{fontenc}
\usepackage[perpage]{footmisc}
\usepackage[left=2cm,right=2cm,top=2cm,bottom=2cm]{geometry}
\usepackage{graphicx,eso-pic}
\usepackage[bookmarksdepth=4]{hyperref}
\usepackage[hyperpageref]{backref}  
\usepackage[utf8]{inputenc}
\usepackage{lmodern}
\usepackage{mathptmx}
\usepackage[most]{tcolorbox}
\usepackage{setspace}
\usepackage{titling}
\usepackage{tikz} 	
\usepackage{transparent}
\usepackage[parfill]{parskip}
\usepackage{wallpaper}
\usepackage{microtype}
\usepackage{libertine}
\usepackage{parskip}
\usepackage{float}
\usepackage{indentfirst}

\definecolor{DARKRED}{rgb}{0.545,0.0,0.0}
\definecolor{DARKGREEN}{rgb}{0.0078,0.4314,0.5059}
\definecolor{LIGHTGREEN}{rgb}{0,0.6706,0.7412}
\definecolor{LIGHTBLUE}{rgb}{0,0.6,0.8667}
\definecolor{LIGHTORANGE}{rgb}{1,0.6,0.20}
\definecolor{LIGHTGRAY}{rgb}{0.6314,0.7804,0.8784}

\hypersetup{
    colorlinks=true,                          
    linkcolor=LIGHTBLUE, 
    citecolor=DARKGREEN, 
    urlcolor=blue,
    linktoc=all
}

\graphicspath{figures}
\DeclareGraphicsExtensions{.jpg,.mps,.pdf,.png,.eps} 

\newcommand{\via}{\textit{via }}
\newcommand{\etc}{\textit{etc }}
\newcommand{\eg}{\textit{eg }}
\newcommand{\cf}{\textit{cf }}

\newcommand{\degre}{$^\circ\mathrm{C}$ }

\title{\huge On the mechanism of thermal self-regulation of trees: a kind of homeothermic observation}
\author{
\normalsize
    \begin{minipage}[t]{\textwidth}
        \centering
        Boulé Jean-Baptiste\textsuperscript{1}, de Bremond d'Ars Jean\textsuperscript{2}, Courtillot Vincent\textsuperscript{3},\\
        Gèze Marc\textsuperscript{4}, Gibert Dominique\textsuperscript{5}, Le Mouël Jean-Louis\textsuperscript{3},\\
        Lopes Fernando\textsuperscript{1}, Maineult Alexis\textsuperscript{6}, Zuddas Pierpaolo\textsuperscript{7}
    \end{minipage} \\
    \ \\
    \footnotesize\textsuperscript{1}Museum National d’Histoire Naturelle, CNRS UMR 7196, Sorbonne Université, Paris, France\\
    \footnotesize\textsuperscript{2}Université de Rennes, CNRS  UMR 6118, Géosciences Rennes, Rennes, France\\
    \footnotesize\textsuperscript{3}Académie des Sciences, Institut de France, Paris, France\\
    \footnotesize\textsuperscript{4}Museum National d’Histoire Naturelle, CEMIM, Sorbonne Université, Paris, France\\
    \footnotesize\textsuperscript{5}Université de Lyon, ENSL, CNRS UMR 5276, Villeurbanne, France\\
    \footnotesize\textsuperscript{6}Laboratoire de Géologie de l'ENS, UMR 8538, Paris, France\\
    \footnotesize\textsuperscript{7}Sorbonne Université, CNRS UMR7619, Paris, France
}

\date{}

\begin{document}
\maketitle

\begin{abstract}
	What is certain is that surface temperatures around the globe vary considerably, regardless of the time scales or underlying causes. Since 1850, we have observed an average increase in global surface temperature anomalies of 1.2\degre and a median increase of 0.7\degre: this overall difference masks significant regional differences. Nearly 60\% of the world's population now lives in urban areas, where vegetation cover has been significantly reduced, despite the paradoxical fact that vegetation plays an important role in regulating the thermal environment (\eg through the shading provided by tree canopies). Continuous electrical and thermal measurements of trees in a Parisian grove (France) show and quantify that canopies are not the only protectors against heat waves; we must also consider the role of tree trunks. It is clear that these trunks probably regulate themselves, possibly by modulating the uptake of groundwater, whose geothermal stability is well established at a depth of just one metre. This quantitative observation should not be overlooked in the urban planning of our cities.
\end{abstract}

\section{Introduction} 
	
	The climatic cycles and their vicissitudes, always variable throughout time, are currently leading us to an increase in the anomaly of ground-measured temperature by $\sim$1.2\degre on average (\cf \shortciteNP{masson2021}), and by $\sim$0.7\degre in median (\cf \shortciteNP{courtillot2023a}). This climate variability, both spatially and temporally, means that on the one hand, we do not find the same tree species at all latitudes (\eg \shortciteNP{randin2013,boisvert2014}), with conifers being primarily present in colder areas and deciduous trees in more temperate zones. On the other hand, for the same species, the number of individuals in a homogeneous forest and their vitality can vary over time (\eg \shortciteNP{svenning2008,fang2018,courtillot2023b}). For some time now, we have been observing a reorganization of species, or rather an adaptation (forest size, individual sizes, \etc), occurring with increasing temperature (\eg \shortciteNP{saxe2001,trumbore2015}), increasing  primarily observed in the high latitudes of the Northern Hemisphere (\cf \shortciteNP{balting2021}). 

	We already know, or at least we have strong reasons to believe, that tree canopies, and forest canopies in particular, play a crucial role in climate interactions (\cf \shortciteNP{bonan2008,frey2016,zellweger2020}). Through photosynthesis (\cf \shortciteNP{einstein1912}, which is an endothermic chemical reaction using solar energy to convert carbon dioxide and water into glucose and oxygen, tree leaves contribute, to some extent, to regulating heat at the Earth's surface. It is not surprising, then, that in addition to photosynthesis, if we consider the purely filtering aspect of solar radiation by tree canopies (which can reduce up to 100 W.m$^{-2}$), micro-climates can be found within these forests (\eg \shortciteNP{grimmond2000,holst2004,vonarx2012,dodorico2013,defrenne2021,gril2023}). The question of adaptation to temperature fluctuations in plants in general, and trees in particular, adaptation in the sense of Lamarck, is a question worthy of interest and, as we have seen, of vast scientific importance. Yet, there is little literature on the quantitative physiological monitoring of trees when they undergo temperature variations. This is the subject we will address in the continuation of our study.
	
	In a previous study, nearly 20 years old (\cf \shortciteNP{gibert2006}), we demonstrated through continuous electrical measurements that the sap flow circulating in a poplar tree generated a spontaneous electrical signal, known to geophysicists as spontaneous potential (SP, \eg \shortciteNP{journiaux2009}), through a well-known physical phenomenon: electrokinetics. On this occasion, we demonstrated the existence of a diurnal electrical oscillation, albeit greatly attenuated in amplitude, nevertheless present in winter. Thus, we showed that sap flowed even in winter. In a more recent study (\cf \shortciteNP{lemouel2024}), where we revisited the data from \shortcite{gibert2006}, we showed that regardless of the position on the poplar tree (roots, trunk, branches) where the measurements were made, the electrical signals were more or less decomposable into sums of pseudo-cycles, with nominal periods corresponding to those of the main lunar-solar tides. The sum of these pseudo-cycles represented over 70\% of the variability of the electrical signal. We thus highlighted a general mechanism for sap circulation; this electrokinetic and physiological signal of the poplar tree was thus the living counterpart to harmonic pumping in geosciences (\eg \shortciteNP{maineult2008,allegre2014}). Motivated by these results, since 2018, we have built a geophysical observatory of the living in the center of Paris (France), within the Jardin des Plantes of the Muséum National d'Histoire Naturelle (MNHN). Over a dozen trees are continuously monitored and in real-time, using sap flow sensors (Granier probe), SP measurements (with simples stainless steel electrodes), and also, what interests us here, temperature measurements \via platinum probes (Pt-100 and Pt-1000).

	In Section \ref{sec02}, the next section, we will present the site and the acquisition system, and to illustrate our points, we will present some data. In Section \ref{sec03}, we will discuss temperature in trees proper. First, we will see how it remains, on average, incredibly stable throughout the seasons, while exhibiting surprising behaviors. Then, we will discuss the relationship between sap flow and temperature within this thermally isolating cylinder that can be a tree. Finally in Section \ref{sec04}, we will conclude in light of these results.

\section{Site, Acquisition Equipment and data Presentation\label{sec02}}

\begin{figure}[H]
	\begin{center}
		\includegraphics[width=16cm]{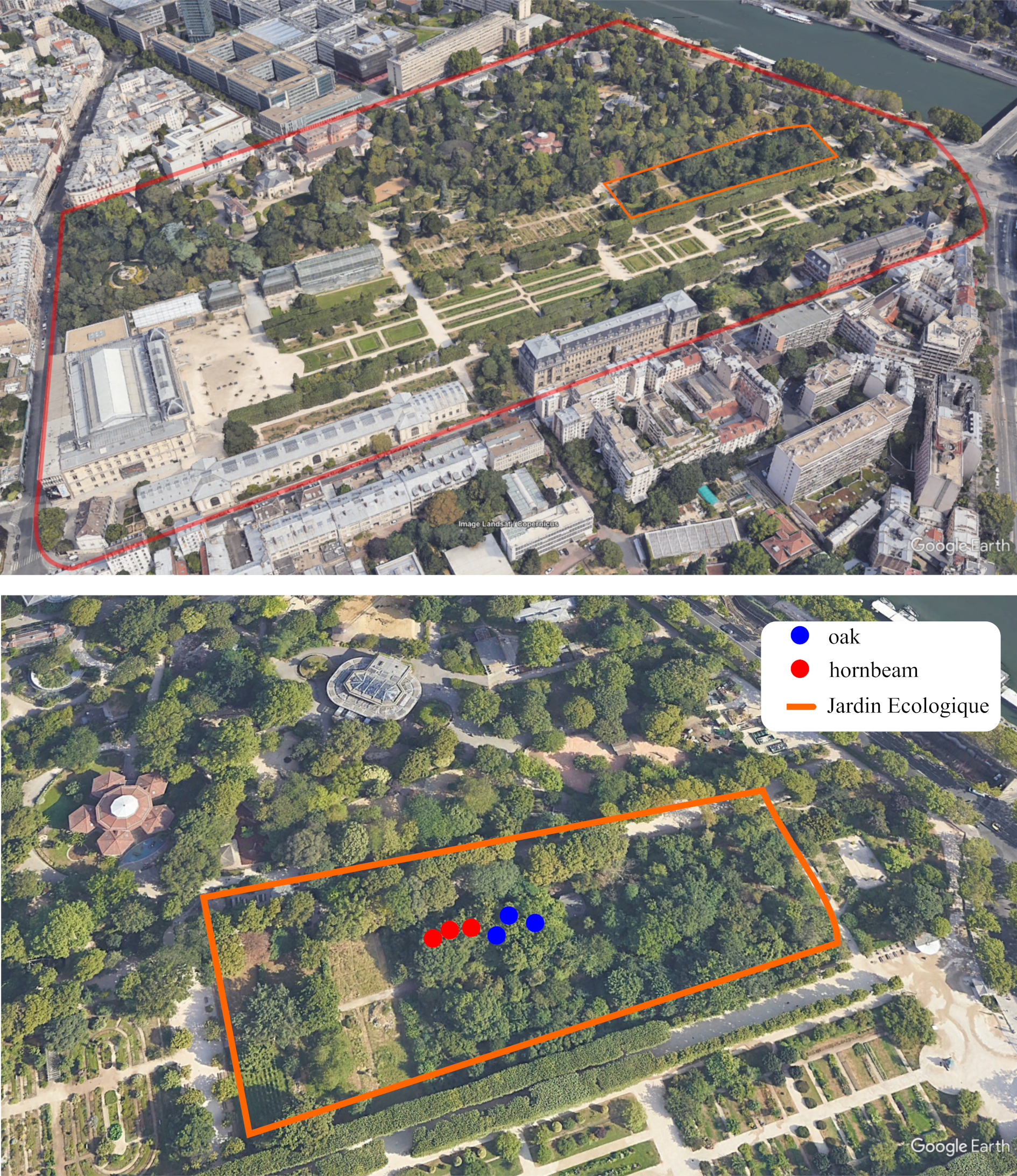}
	\end{center}
	\caption{ At the top, outlined in red, the ‘Muséum National d’Histoire Naturelle’ (Paris,
France) and its ‘Jardin des Plantes’; outlined in orange, the ‘Jardin Écologique’. At the bottom, an
enlargement, the red and blue dots respectively mark the positions of the 3 hornbeams and the 3
oaks.\label{fig:01}}
\end{figure}

	The observation site, located in the center of Paris (France), is part of the Muséum National d'Histoire Naturelle (\cf Figure \ref{fig:01} spotted by the red shape). It is an ecological garden (\cf Figure \ref{fig:01} spotted by the orange shape), which contains several dozen trees of various species. This ecological garden is not open to the public.
	
	For this study, we will present and discuss measurements (electrical and thermal) taken on 3 oak trees and 3 horn-beams, as well as temperature measurements in the soil and air. Regardless of the tree, we systematically placed a crown of 4 electrodes around the trunk at 150 cm above the ground, with each electrode oriented towards one of the cardinal points. The electrodes are simple stainless steel rods. Also, for each tree, we measure temperatures on the north-facing side of the trunk at 50 and 100 cm above the ground; between these two points, we also measure the temperature outside the trunk and at a height of 75 cm above the ground. The temperature probes used are Pt-100. The first hornbeam is equipped with an ICT sap flow sensor. All measurements have been acquired and sampled at one second intervals using Gantner modular acquisition systems. These data acquisition have significantly better performance than the Keitley 2701 used in by \shortciteN{gibert2006}: a high impedance ($ > $ 100 M$\Omega$) and a dynamic range of 24 bits capable of sampling up to 20 kHz per channel.

\begin{figure}[H]
	\begin{center}
		\includegraphics[width=10cm]{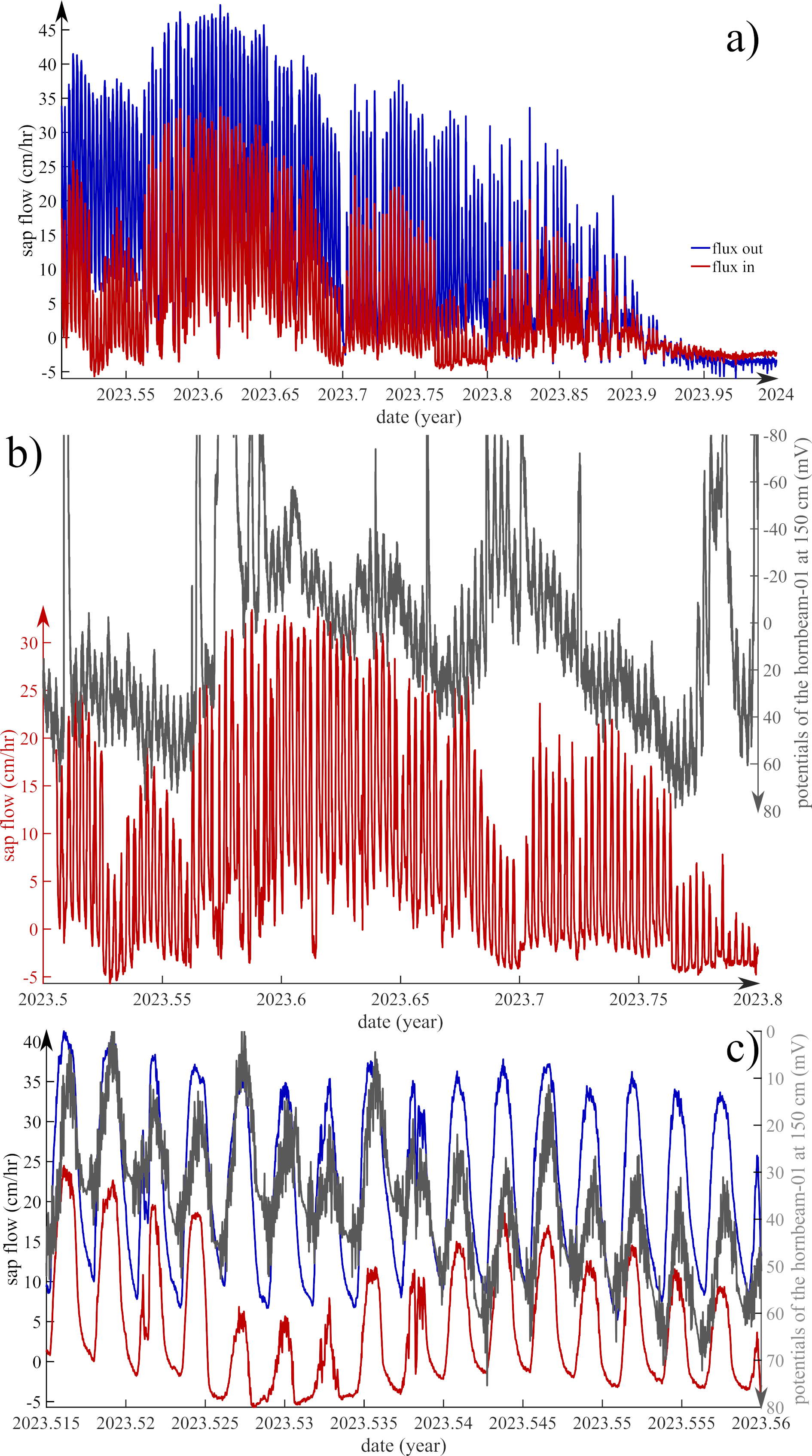}
	\end{center}
	\caption{Validation of electrical measurements by comparison with those from a sap flow sensor. a) Incoming and outgoing flux from the Granier sensor, b) overlay between incoming flux and the electrical potential measured next to the Granier probe, c) overlay of all signals to appreciate the phase evolution of the diurnal oscillation.\label{fig:02}}
\end{figure}

	Each of our trees, the 3 oaks and the 3 horn-beams, will be identified by a number. For example, in Figures \ref{fig:02}, we present the sap flow measurements and the electrode closest to the Granier probe on hornbeam 01. This probe was placed for convenience reasons on the southwest side of hornbeam 01, about 180\ cm from the ground. The closest electrode, oriented due south, was at 150\ cm. In Figure \ref{fig:02}a, from June 01, 2023, to December 31, 2023, red represents the upward flow entering the Granier probe, and blue represents the upward flow exiting the said Granier probe. In Figure \ref{fig:02}b, from June 01, 2023, to September 01, 2023, gray represents the electrical potential measured on the south electrode of Hornbeam 01; we have reversed the ordinate axis. Red represents the sap flow measured during the same period. It is typical to observe polarization peaks or sudden and transient changes in amplitudes in electrical measurements; we observe these particularly in May. Nevertheless, it seems that the trend of the potential curve (gray) closely follows the envelope of the incoming sap flow (red). In Figure \ref{fig:02}c, we present a new enlargement and a superposition of all the signals just mentioned, between May 11, 2023, and May 27, 2023. Here too, we have inverted the axis of electrical potentials. The diurnal cycles we observe, both in sap flow and in electrical potentials, are indeed in phase, or rather in opposition of phase (the potential axis having been reversed).

	Undoubtedly, the main pseudo-oscillation measured in trees, the diurnal oscillation (an earth tide), is evident in both the electrical and the sap flow signals, and thus, as \shortciteN{gibert2006} showed more than 20 years ago, an electrical signal attributable to a complex phenomenon of charge displacement (but not only, \eg \shortciteNP{maineult2005}) is indeed associated with sap flow: an electrokinetic phenomenon. We can therefore have confidence in our potential measurements.
	
	The acquisition system (the nature of the measurements, the number of sensors, \etc) as well as the number of trees we monitor have been constantly increasing since 2019. The subject we address in this paper is the thermal regulation of trees, and specifically, in our living observatory (\textbf{MNHN}), the longest time series combining both electrical and thermal measurements begins in January 2023 and concerns oak 01. In Figure 03, we present a year of nearly continuous electrical and thermal acquisitions conducted on oak 01 and in its immediate vicinity (subsoil + ambient air). What immediately strikes us upon viewing this figure is that the temperature trends (at the top) and the potential trends (at the bottom) evolve in a mirror-like fashion over time: the temperature trends increase by approximately 8-10\degre from January 2023 to almost 18-20\degre in June of the same year, then decrease to around 8-10\degre by December 2023. Conversely, the electrical potentials decrease from an average of about 1200 mV for the four electrodes (which is significant). They reach a minimum average of around 500 mV in June 2023, then increase again but do not reach their initial potentials by December of the same year. Clearly, the data indicate a link between sap flow, which we have mentioned is related to electrical potential, and thermal regulation in the tree. This link is evidently not linear but appears to be more complex.

\begin{figure}[H]
	\begin{center}
		\includegraphics[width=16cm]{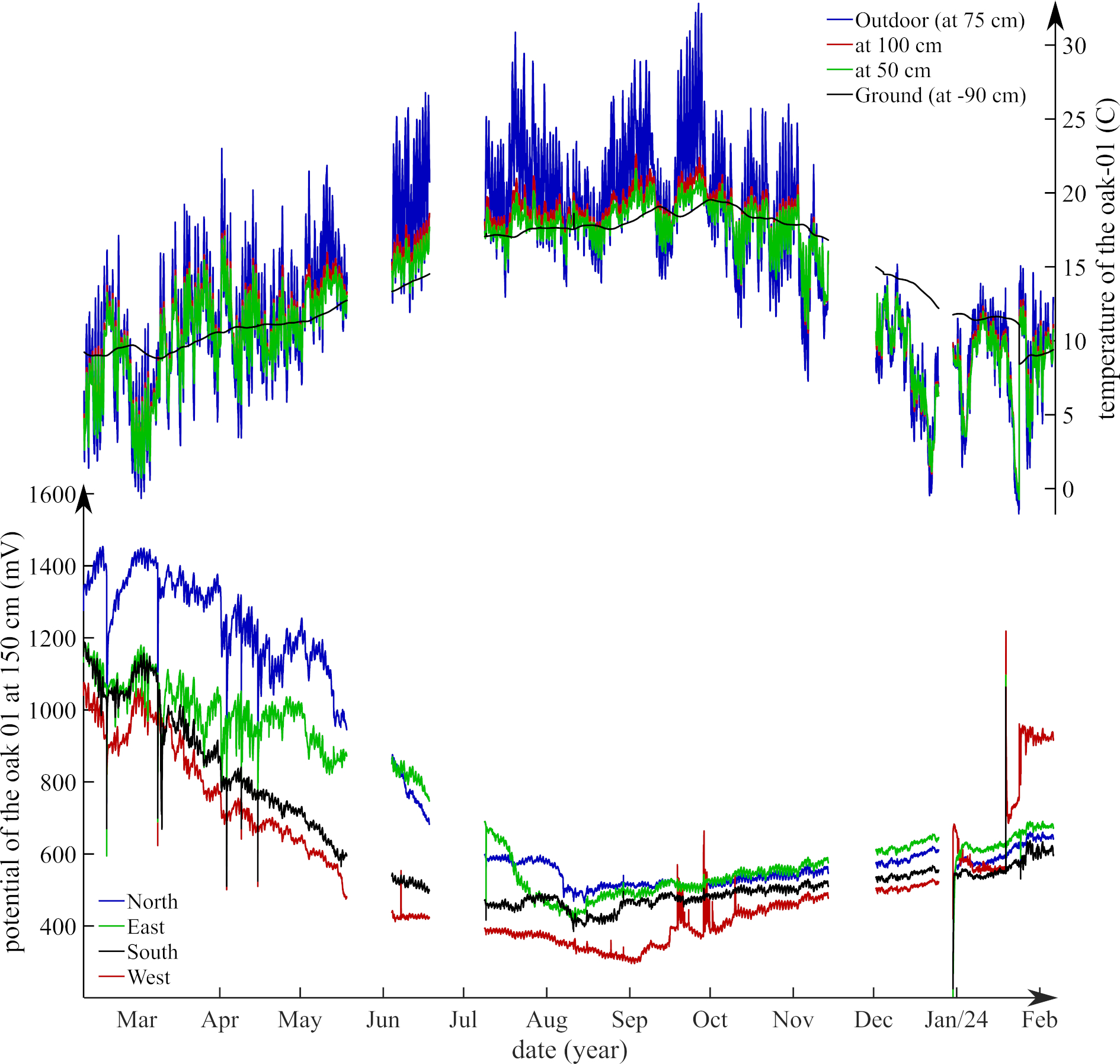}
	\end{center}
	\caption{Temporal evolution since 2023: at the top, temperatures recorded within oak 01 at 50 cm and 100 cm (red curve), as well as in the soil at 90 cm depth (black curve), and outside the tree (attached to the tree) at 75 cm (blue curve); at the bottom, electrical potentials measured  at 150 cm on the North direction in blue, on the East direction in green, on the South in black, and on the West in red.\label{fig:03}}
\end{figure}

\section{Study of the longest time series of temperature, that of oak tree 01\label{sec03}}
  For all tree species that we monitor, we systematically bury a temperature probe at a depth of over 90 cm in the soil. The soil and the nearby surface, from which trees extract water and nutrients, obey the laws of heat conduction (\cf \shortciteNP{fourier1888}), also known as geothermal laws. In its simplest expression, temperature (\degre) obeys the following conduction equation,
\begin{equation}
	\dfrac{\partial ^{2} T}{\partial^{2}z } - \dfrac{1}{\chi} \dfrac{\partial T}{\partial t} = 0,
	\label{eq:01}
\end{equation}  
  
 equation (\ref{eq:01}) in which $z$ is the direction perpendicular to the ground indicating in our case the depth (in meters), $\chi$ is the thermal diffusivity of the soil (in cm$^{2}$.s$^{-1}$), and $t$ is time. Suppose only variations in external temperature drive the temperature in the soil, then an immediate solution of equation (\ref{eq:01}) is,  
\begin{equation}
	T = T_{0} \exp(-z\sqrt{\omega/2\chi}) \cos(\omega.t - z\sqrt{\omega/2\chi}).  
  	\label{eq:02}
\end{equation}  

	This solution (\ref{eq:02}) corresponds to a heat wave with periodicity $\omega$ attenuating with depth according to the term in the exponential function. This heat wave, with a wavelength of $\lambda = 2\pi(\omega/2\chi)^{-1/2}$, attenuates by a factor of $\exp(-2\pi) \approx 10^{-3}$ or a soil whose thermal diffusivity is between 0.02 (dry) and 0.033 (moist) cm$^{2}$.s$^{-1}$ at a depth ranging from 0.9 to 1 meter for the diurnal oscillation. This is the reason why we buried our PT-100 sensors at around 90 cm, and it's the reason why the temperature in the soil (\cf black curve Figure \ref{fig:03}a) doesn't exhibit any diurnal oscillation. This relationship (\ref{eq:01}), which is a diffusion relationship, also informs us, to a certain extent (depth), that the temperature in the soil, while filtered and smoothed from its abrupt variations, also indicates that heat waves of longer periods associated with climates and paleoclimates, for example, take some time to penetrate the said soil. In other words, the temperature in the soil presented by the black curve in Figure \ref{fig:03}a is not a consequence of the blue outdoor temperature curve; it is the memory of a past temperature variation that, at this depth, corresponds to an average annual oscillation, the seasonal oscillation.

\begin{figure}[H]
	\begin{center}
		\includegraphics[width=16cm]{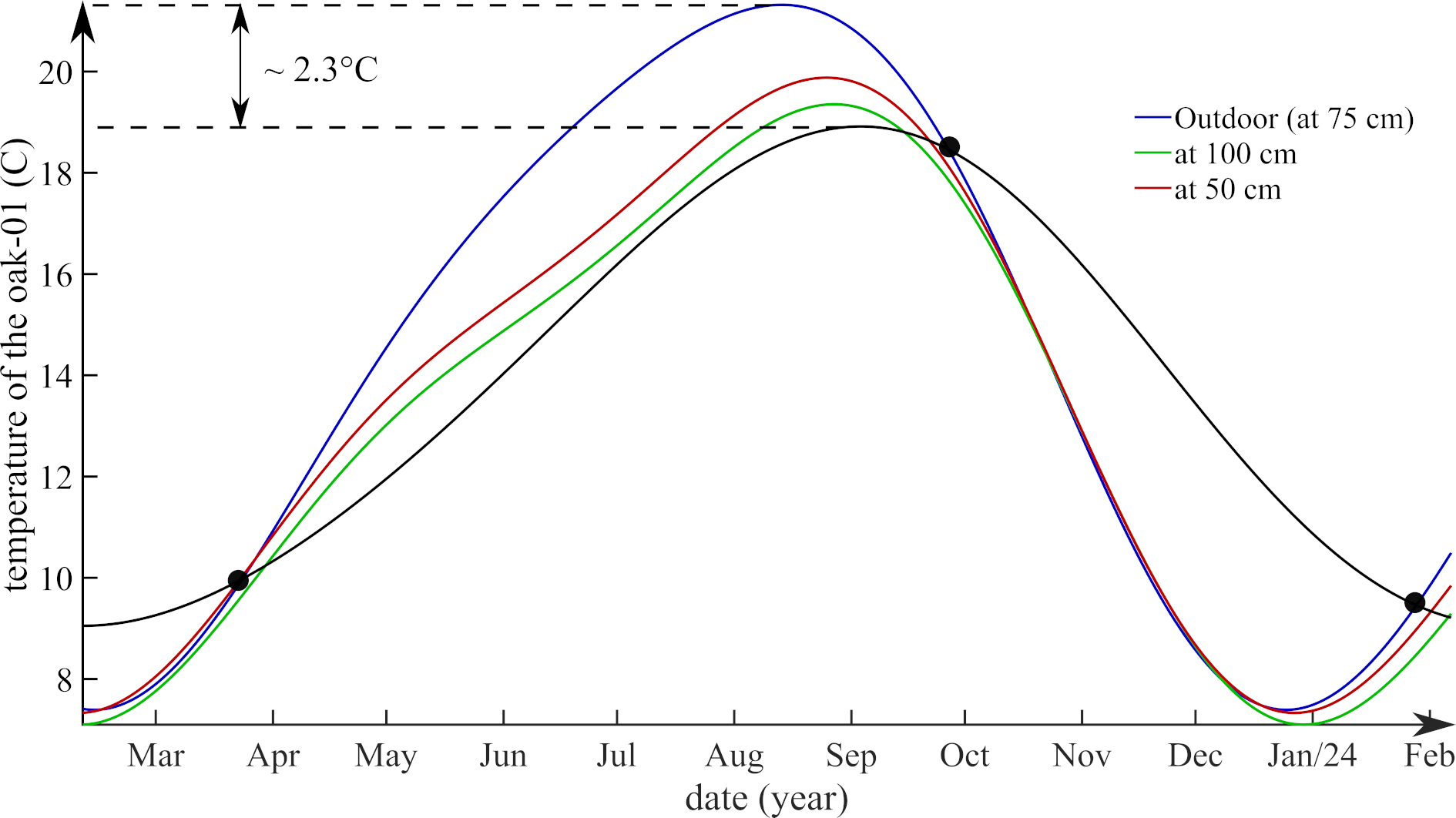}
	\end{center}
	\caption{Estimated medians of the temperature curves presented in Figure \ref{fig:03}a. Data gaps have been extrapolated.\label{fig:04}}
\end{figure}

  In Figure \ref{fig:04}, we calculated the median temperature curves measured inside and outside oak 01. A remarkable behavior unfolds before us. Throughout one year, from February 2023 to February 2024, the median temperature curves outside the oak (in blue) and within the soil (in black) only intersect 3 times: at the end of March 2023, the beginning of October 2023, and the end of January 2024. We marked these dates with black dots. When the temperature within the soil is warmer than the temperature outside the tree, from January to the end of March 2023 and between the beginning of October 2023 and the end of January 2024, the median temperatures inside the tree (in green and red) almost perfectly follow, if not entirely overlap with, the temperature outside the oak. The behavior is completely different between April and October 2023, the warmest period in our latitudes, during which, on one hand, the median temperature within the tree is no longer perfectly superimpose, and on the other hand, they tend to approach the median temperature of the soil. At the peak of the summer period, concerning trends, we can observe a difference of over 2\degre between the tree and its environment. However, as we will see later, this difference will be much greater concerning the diurnal oscillation.
  
\begin{figure}[H]
	\begin{center}
		\includegraphics[width=16cm]{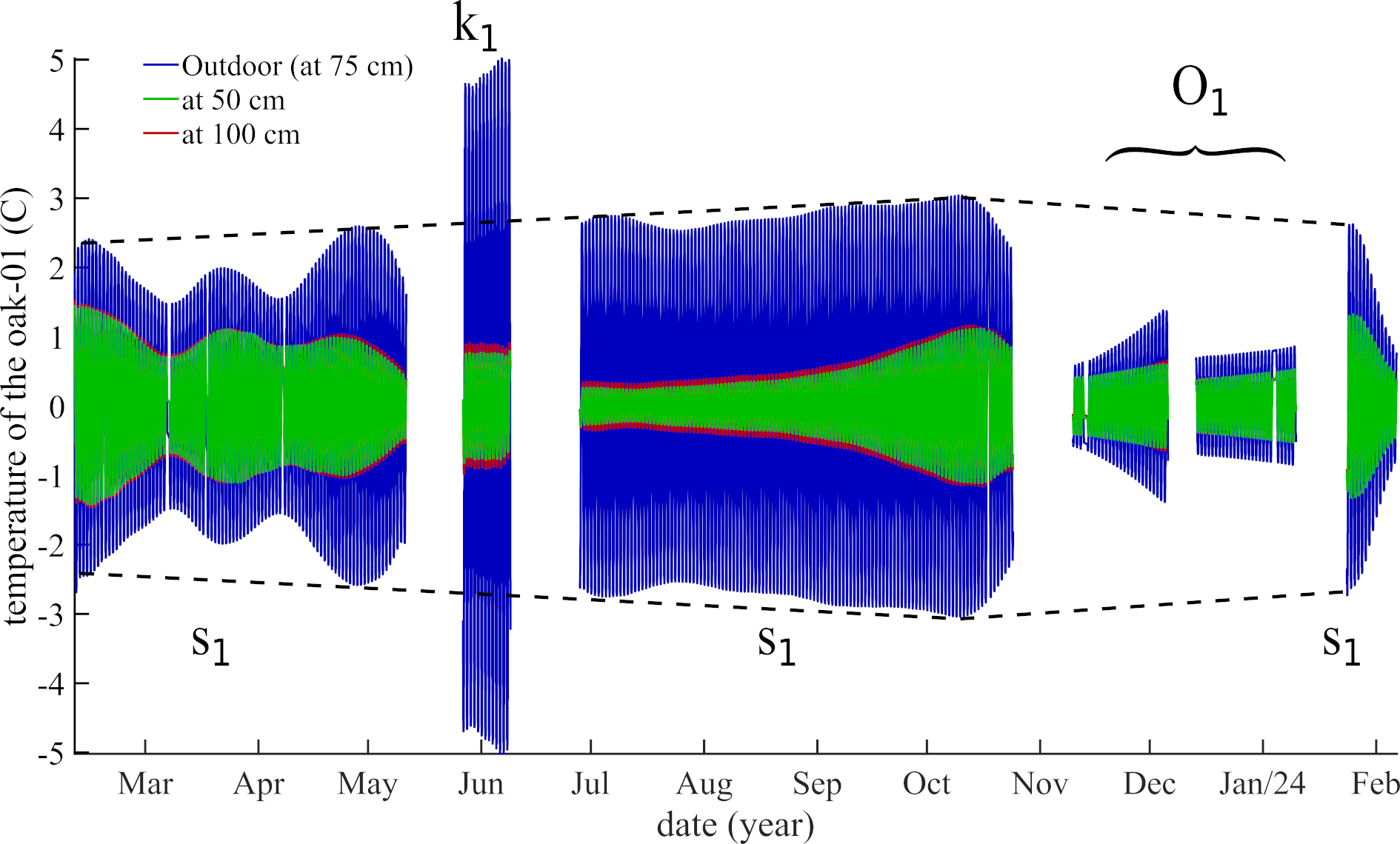}
	\end{center}
	\caption{Diurnal pseudo-cycles extracted by SSA from temperature data in Figure \ref{fig:03}a. Four components linked to lunar-solar tides are clearly visible both inside and outside the tree. The black dashed line represents the theoretical envelope of the S1 tide.\label{fig:05a}}
\end{figure}

\begin{figure}[H]
	\begin{center}
		\includegraphics[width=16cm]{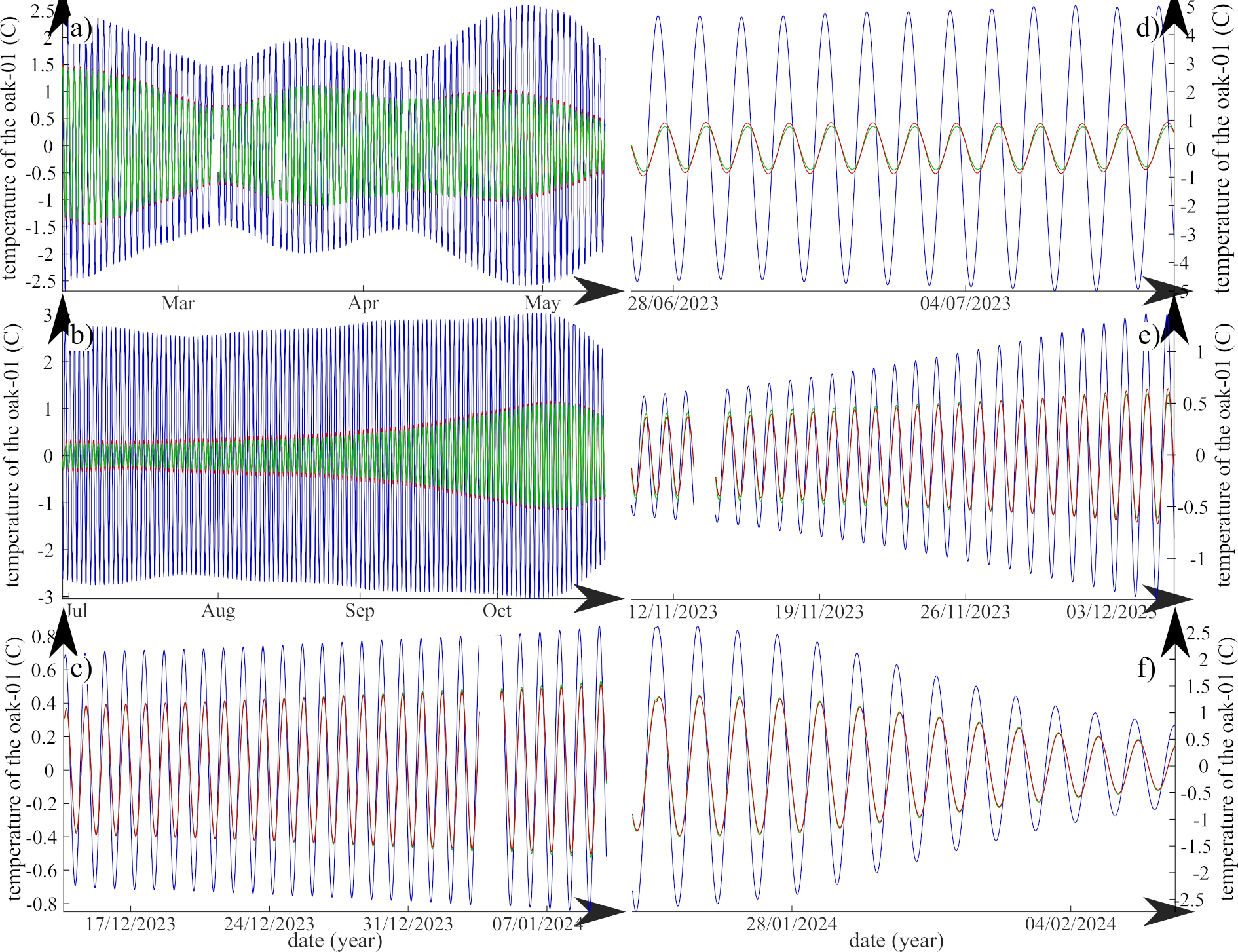}
	\end{center}
	\caption{Each of the diurnal pseudo-cycles from Figure \ref{fig:05b} is presented separately.\label{fig:05b}}
\end{figure}

  In Figures \ref{fig:05a} and \ref{fig:05b}, we have overlaid the pseudo-diurnal oscillations extracted using the Singular Spectrum Analysis (SSA) method. SSA is capable of extracting from a time series (continuous or not), if they exist, a trend as well as all the non-stationary pseudo-cycles that compose the said series. SSA takes advantage of the properties of descending order diagonal matrices (Hankel/ Toeplitz matrices, \cf \shortciteNP{lemmerling2001}) and their orthogonalization by singular value decomposition (SVD, \cf \shortciteNP{golub1971}). We invite readers to explore the book on the subject of \shortciteN{golyandina2013}, or for a detailed summary of the method, to read the methodological section of \shortciteN{lopes2022}.
  
  One striking characteristic that we have already illustrated when commenting on Figures \ref{fig:02} and \ref{fig:03} is that an electrical signal is associated with sap movement (\eg \shortciteNP{gibert2006}), and thus with a temperature variation. On the occasion of \shortciteN{lemouel2024}, we showed that the electrical signals measured in 2003 in the poplar tree of Remungol were 70\% the sum of the main Earth tides The problem with Earth tides is that they are found absolutely everywhere because the forcing, for the shortest periods, lunar-solar, drives several geophysical phenomena. Thus, we can find them in variations in the length of  day (and therefore in insolation variations; \eg \shortciteNP{ray2014,lemouel2019}) as well as in volcanic eruptions, for example (\eg \shortciteNP{dumont2020,lemouel2023}). The temperature measurements inside and outside the tree are no exception.

	In Figure \ref{fig:05a}, we simply aligned semi-diurnal pseudo-cycles extracted by SSA over time. There are six main lunar-solar tides around 24 hours. Here, we have detected three within the tree. Firstly, the primary solar tide (S1), with a period precisely of 1 day. We identified it between February 1st and May 11th, 2023, during this period the temperature difference between my external and internal diurnal oscillations within the tree appears constant, around 1\degre. S1 is also identified between June 28th and October 24th, 2023; here we observe a temperature difference exceeding 3\degre. We find this tide once again starting from January 24th, 2024, with a temperature difference of less than 1\degre. For these three time intervals, we have the same periodicities both outside and inside the tree, which are: 1.008$\pm$0.005 days, 1.003$\pm$0.005 days, and 1.001$\pm$0.004 days. Outside of these time intervals, surprisingly, two other lunar-solar tides are expressed, tides which incidentally are the most significant. Firstly, the tide (K1), linked to the syzygy of the Moon and the Sun, with a theoretical period of 0.990 days and observed between May 27th and June 8th, 2023, both inside and outside the oak at 0.999$\pm$0.004 days. The second recorded tide, purely lunar (O1), has a theoretical period of 1.075 days and was identified within the oak from November 10th, 2023, to January 9th, 2024, at 1.057$\pm$0.023 days. With regards to the (K1) tide, at the height of summer, we observe a temperature amplitude difference of over 5\degre between the external and internal diurnal variations within the tree. In Figure \ref{fig:05b}, to better appreciate the measurements, we have separated the curves into 6 temporal blocks.
	
\section{Discussion\label{sec04}}
	In order to better understand the interaction between climate and forests in terms of temperature exchanges, we did not want to limit ourselves to the canopy, for which there is a multitude of papers in the literature on the subject, but rather to the physiology of an entire tree. Within the Ecological Garden of the MNHN, for over 3 years, we have been building an observatory of life and geophysics in the center of Paris (\cf Figure \ref{fig:01}). 
	
	Trees can be schematized by vertical cylinders through which fluids circulate pumped from the subsoil, thus a priori at a constant temperature on the meteorological scale. These fluid movements generate an electrical signal through an electrokinetic mechanism, whose main oscillation is the diurnal tide (\cf \shortciteNP{gibert2006}). We have definitively verified this assertion by comparing, on a hornbeam tree using a \shortciteN{granier1985} probe, the variations in sap flow and those of the electrical signal SP (\cf Figure \ref{fig:02}). On one hand, the trend of the electrical potential perfectly follows the envelope of the measured sap flow (\cf Figure \ref{fig:02}b); on the other hand, the diurnal oscillations of the two physical phenomena are in phase, or rather in constant phase opposition (\cf Figure \ref{fig:02}c). This electrokinetic signal is reminiscent of geophysical phenomena such as harmonic pumping (\cf \shortciteNP{maineult2008}), which, through lunar-solar tides, causes deep oscillations in groundwater. This celestial harmonic pumping phenomenon is a general phenomenon that applies to our entire planet, regardless of the scales of time and space. Therefore, it was not surprising that, after reanalyzing the electric potential data on the Poplar tree from \shortciteN{gibert2006}, we found almost all the main terrestrial tides in the electrokinetic signal, and thus in the sap flow (\cf \shortciteNP{lemouel2024}). It was therefore not surprising that we found the same terrestrial tides, but this time linked to insolation, in the temperature variations of our trees. 
	
	The longest time series we have at the MNHN, combining both electric potential and temperature measurements, concerns our oak tree number 01 and begins in January 2023 (\cf Figure \ref{fig:03}). Very clearly, we observe variations in the trends of our electric potentials mirroring those of the temperatures. This result is not really surprising because we had observed that the sap flow and the associated electric signal were in perfect phase opposition (\cf Figure \ref{fig:02}, we had to reverse the axis of the potentials). We estimated the median (thus robust) trends of temperatures outside and inside the tree, as well as in the soil, which highlighted an astonishing phenomenon of thermal self-regulation of the oak tree (\cf Figure \ref{fig:04}): when the soil temperature is higher than that outside, the oak tree follows the outside temperature; however, when the outside temperature is higher than that of the soil, the temperature curves inside the tree not only no longer overlap but also tend to approach the soil temperature. We observe a median gap of over 2\degre in trends at the hottest point of August 2023. 
	
	In order to refine our observations regarding the thermal self-regulation of the tree, we extracted the diurnal temperature oscillations using Singular Spectrum Analysis. SSA, being a robust signal analysis technique that works well on strictly stationary signals, broadly stationary signals, or non-stationary signals, and also works on discontinuous signals, seemed to be the best choice for us. We then identified, in the tree temperatures, the presence of lunar-solar tides S1, K1, and O1 (\cf Figures \ref{fig:05a} and \ref{fig:05b}) , which are the main diurnal tides. These tides, similar to those detected (\cf \shortciteNP{lemouel2024}) in the electric signals of the Poplar tree from \shortciteN{gibert2006}, are highly precise in terms of the observed periods. Throughout the year, the variations between these diurnal oscillations inside the tree and outside of it can range from less than 1\degre in winter to over 5\degre in the height of summer. This means that, in addition to canopy coverage, there can be over 7\degre difference (2.3\degre from the trend +5\degre from the diurnal oscillation) between the tree and its environment during the hottest periods of the year. The tree thus regulates its temperature, more or less like a homeothermic organism.

	It is important here to discuss a crucial point. As we have already mentioned, a tree can be seen as a vertical cylinder through which a cold fluid circulates, as explained at the beginning of section (\ref{sec03}). Therefore, there are calories and frigories exchanged between this cylinder and its environment. In order to evaluate, for our oak tree 01, the quantity of heat exchanged, we can, in a very simplified perspective, apply the following Fourier's law,
\begin{equation}
	\mathcal{Q} = \kappa *A *\dfrac{\Delta T}{d}
	\label{eq:03}
\end{equation}	

	This very simplistic law involves $\mathcal{Q}$ the amount of transferred heat (W), $\kappa$ the thermal conductivity of wood which is between 0.2 and 0.4 W.m$^{-1}$.\degre$^{-1}$, $\Delta T$ the temperature difference (\degre) between the cylinder and the exterior, the surface area for heat transmission (m$^{2}$), and the height of the cylinder (m). A simple order of magnitude calculation gives us, for a perfectly cylindrical tree, with a height of 20 m, a perimeter of 2 m, a thermal conductivity of 0.3 W.m$^{-1}$.\degre$^{-1}$, at a temperature of 15\degre in an environment at 25\degre (a delta of 10\degre), a theoretical heat exchange of 120W. This value is perfectly compatible with those found in the literature, measured for example by \shortciteN{dossantos2008}.

	Another surprising aspect is that it is known that wood is the best thermal insulator. However, as can be seen in curves 'd', 'e', and 'f' of Figure \ref{fig:05b}, the diurnal oscillations within the tree, not present in the soil, can be in phase quadrature with the diurnal oscillations of the environment (curves 'd' and 'e') or perfectly in phase in winter (curve 'f'). Therefore, the diurnal heat wave takes between 0 and 6 hours to warm our trees, which is very rapid and paradoxical with the notion of thermal insulation that we just mentioned.

	In light of these observations, we believe it is important, before studying the interaction between a forest and the climate, or before studying the ability of a tree to cool its environment, to better understand and study the physiology and thermal regulation of trees. Clearly, even without considering their canopies, trees are involved in variable  (as trees self-regulate) transfers of hundreds of watts of heat/cold.
\bibliographystyle{fchicago}
\bibliography{biblio}

\begin{thebibliography}{}

\bibitem[\protect\citeauthoryear{Allegre, Maineult, Lehmann, Lopes, et
  Zamora}{Allegre {\em et~al.}}{2014}]{allegre2014}
Allegre, V., A.~Maineult, F.~Lehmann, F.~Lopes, et M.~Zamora (2014).
\newblock Self-potential response to drainage--imbibition cycles.
\newblock {\em Geophysical Journal International\/}~{\em 197\/}(3), 1410--1424.

\bibitem[\protect\citeauthoryear{Balting, AghaKouchak, Lohmann, et
  Ionita}{Balting {\em et~al.}}{2021}]{balting2021}
Balting, D.~F., A.~AghaKouchak, G.~Lohmann, et M.~Ionita (2021).
\newblock Northern hemisphere drought risk in a warming climate.
\newblock {\em NPJ Climate and Atmospheric Science\/}~{\em 4\/}(1), 61.

\bibitem[\protect\citeauthoryear{Boisvert-Marsh, P{\'e}ri{\'e}, et
  de~Blois}{Boisvert-Marsh {\em et~al.}}{2014}]{boisvert2014}
Boisvert-Marsh, L., C.~P{\'e}ri{\'e}, et S.~de~Blois (2014).
\newblock Shifting with climate? evidence for recent changes in tree species
  distribution at high latitudes.
\newblock {\em Ecosphere\/}~{\em 5\/}(7), 1--33.

\bibitem[\protect\citeauthoryear{Bonan}{Bonan}{2008}]{bonan2008}
Bonan, G.~B. (2008).
\newblock Forests and climate change: forcings, feedbacks, and the climate
  benefits of forests.
\newblock {\em science\/}~{\em 320\/}(5882), 1444--1449.

\bibitem[\protect\citeauthoryear{Courtillot, Boul{\'e}, Le~Mou{\"e}l, Gibert,
  Zuddas, Maineult, G{\`e}ze, et Lopes}{Courtillot {\em
  et~al.}}{2023}]{courtillot2023b}
Courtillot, V., J.-B. Boul{\'e}, J.-L. Le~Mou{\"e}l, D.~Gibert, P.~Zuddas,
  A.~Maineult, M.~G{\`e}ze, et F.~Lopes (2023).
\newblock A living forest of tibetan juniper trees as a new kind of
  astronomical and geophysical observatory.
\newblock {\em arXiv e-prints\/}, arXiv--2306.

\bibitem[\protect\citeauthoryear{Courtillot, Lopes, Gibert, Boul{\'e}, et
  Le~Mou{\"e}l}{Courtillot {\em et~al.}}{2023}]{courtillot2023a}
Courtillot, V., F.~Lopes, D.~Gibert, J.~Boul{\'e}, et J.~Le~Mou{\"e}l (2023).
\newblock On variations of global mean surface temperature: When laplace meets
  milankovi$\backslash$'c.
\newblock {\em arXiv preprint arXiv:2306.03442\/}.

\bibitem[\protect\citeauthoryear{De~Frenne, Lenoir, Luoto, Scheffers,
  Zellweger, Aalto, Ashcroft, Christiansen, Decocq, De~Pauw, {\em
  et~al.}}{De~Frenne {\em et~al.}}{2021}]{defrenne2021}
De~Frenne, P., J.~Lenoir, M.~Luoto, B.~R. Scheffers, F.~Zellweger, J.~Aalto,
  M.~B. Ashcroft, D.~M. Christiansen, G.~Decocq, K.~De~Pauw, {\em et~al.}
  (2021).
\newblock Forest microclimates and climate change: Importance, drivers and
  future research agenda.
\newblock {\em Global Change Biology\/}~{\em 27\/}(11), 2279--2297.

\bibitem[\protect\citeauthoryear{D'odorico, He, Collins, De~Wekker, Engel, et
  Fuentes}{D'odorico {\em et~al.}}{2013}]{dodorico2013}
D'odorico, P., Y.~He, S.~Collins, S.~F. De~Wekker, V.~Engel, et J.~D. Fuentes
  (2013).
\newblock Vegetation--microclimate feedbacks in woodland--grassland ecotones.
\newblock {\em Global Ecology and Biogeography\/}~{\em 22\/}(4), 364--379.

\bibitem[\protect\citeauthoryear{dos Santos~Michiles et Gielow}{dos
  Santos~Michiles et Gielow}{2008}]{dossantos2008}
dos Santos~Michiles, A.~A. et R.~Gielow (2008).
\newblock Above-ground thermal energy storage rates, trunk heat fluxes and
  surface energy balance in a central amazonian rainforest.
\newblock {\em Agricultural and Forest Meteorology\/}~{\em 148\/}(6-7),
  917--930.

\bibitem[\protect\citeauthoryear{Dumont, Le~Mou{\"e}l, Courtillot, Lopes,
  Sigmundsson, Coppola, Eibl, et Bean}{Dumont {\em et~al.}}{2020}]{dumont2020}
Dumont, S., J.-L. Le~Mou{\"e}l, V.~Courtillot, F.~Lopes, F.~Sigmundsson,
  D.~Coppola, E.~P. Eibl, et C.~J. Bean (2020).
\newblock The dynamics of a long-lasting effusive eruption modulated by earth
  tides.
\newblock {\em Earth and Planetary Science Letters\/}~{\em 536}, 116145.

\bibitem[\protect\citeauthoryear{Einstein}{Einstein}{1912}]{einstein1912}
Einstein, A. (1912).
\newblock Thermodynamische begr{\"u}ndung des photochemischen
  {\"a}quivalentgesetzes.
\newblock {\em Annalen der Physik\/}~{\em 342\/}(4), 832--838.

\bibitem[\protect\citeauthoryear{Fang, Alfaro, et Zhang}{Fang {\em
  et~al.}}{2018}]{fang2018}
Fang, O., R.~I. Alfaro, et Q.-B. Zhang (2018).
\newblock Tree rings reveal a major episode of forest mortality in the late
  18th century on the tibetan plateau.
\newblock {\em Global and Planetary Change\/}~{\em 163}, 44--50.

\bibitem[\protect\citeauthoryear{Fourier}{Fourier}{1888}]{fourier1888}
Fourier, J. B.~J. (1888).
\newblock {\em Th{\'e}orie analytique de la chaleur}, Volume~1.
\newblock Gauthier-Villars.

\bibitem[\protect\citeauthoryear{Frey, Hadley, Johnson, Schulze, Jones, et
  Betts}{Frey {\em et~al.}}{2016}]{frey2016}
Frey, S.~J., A.~S. Hadley, S.~L. Johnson, M.~Schulze, J.~A. Jones, et M.~G.
  Betts (2016).
\newblock Spatial models reveal the microclimatic buffering capacity of
  old-growth forests.
\newblock {\em Science advances\/}~{\em 2\/}(4), e1501392.

\bibitem[\protect\citeauthoryear{Gibert, Le~Mou{\"e}l, Lambs, Nicollin, et
  Perrier}{Gibert {\em et~al.}}{2006}]{gibert2006}
Gibert, D., J.-L. Le~Mou{\"e}l, L.~Lambs, F.~Nicollin, et F.~Perrier (2006).
\newblock Sap flow and daily electric potential variations in a tree trunk.
\newblock {\em Plant Science\/}~{\em 171\/}(5), 572--584.

\bibitem[\protect\citeauthoryear{Golub et Reinsch}{Golub et
  Reinsch}{1971}]{golub1971}
Golub, G.~H. et C.~Reinsch (1971).
\newblock Singular value decomposition and least squares solutions.
\newblock Dans {\em Handbook for Automatic Computation: Volume II: Linear
  Algebra}, pp.\  134--151. Springer.

\bibitem[\protect\citeauthoryear{Golyandina, Korobeynikov, et
  Zhigljavsky}{Golyandina {\em et~al.}}{2013}]{golyandina2013}
Golyandina, N., A.~Korobeynikov, et A.~Zhigljavsky (2013).
\newblock {\em Singular spectrum analysis with R}.
\newblock Springer.

\bibitem[\protect\citeauthoryear{Granier}{Granier}{1985}]{granier1985}
Granier, A. (1985).
\newblock Une nouvelle m{\'e}thode pour la mesure du flux de s{\`e}ve brute
  dans le tronc des arbres.
\newblock Dans {\em Annales des Sciences foresti{\`e}res}, Volume~42, pp.\
  193--200. EDP Sciences.

\bibitem[\protect\citeauthoryear{Gril, Laslier, Gallet-Moron, Durrieu, Spicher,
  Le~Roux, Brasseur, Haesen, Van~Meerbeek, Decocq, {\em et~al.}}{Gril {\em
  et~al.}}{2023}]{gril2023}
Gril, E., M.~Laslier, E.~Gallet-Moron, S.~Durrieu, F.~Spicher, V.~Le~Roux,
  B.~Brasseur, S.~Haesen, K.~Van~Meerbeek, G.~Decocq, {\em et~al.} (2023).
\newblock Using airborne lidar to map forest microclimate temperature buffering
  or amplification.
\newblock {\em Remote Sensing of Environment\/}~{\em 298}, 113820.

\bibitem[\protect\citeauthoryear{Grimmond, Robeson, et Schoof}{Grimmond {\em
  et~al.}}{2000}]{grimmond2000}
Grimmond, C., S.~Robeson, et J.~Schoof (2000).
\newblock Spatial variability of micro-climatic conditions within a
  mid-latitude deciduous forest.
\newblock {\em Climate Research\/}~{\em 15\/}(2), 137--149.

\bibitem[\protect\citeauthoryear{Holst, Hauser, Kirchg{\"a}ssner, Matzarakis,
  Mayer, et Schindler}{Holst {\em et~al.}}{2004}]{holst2004}
Holst, T., S.~Hauser, A.~Kirchg{\"a}ssner, A.~Matzarakis, H.~Mayer, et
  D.~Schindler (2004).
\newblock Measuring and modelling plant area index in beech stands.
\newblock {\em International Journal of Biometeorology\/}~{\em 48}, 192--201.

\bibitem[\protect\citeauthoryear{Jouniaux, Maineult, Naudet, Pessel, et
  Sailhac}{Jouniaux {\em et~al.}}{2009}]{journiaux2009}
Jouniaux, L., A.~Maineult, V.~Naudet, M.~Pessel, et P.~Sailhac (2009).
\newblock Review of self-potential methods in hydrogeophysics.
\newblock {\em Comptes Rendus. G{\'e}oscience\/}~{\em 341\/}(10-11), 928--936.

\bibitem[\protect\citeauthoryear{Le~Mou{\"e}l, Gibert, Courtillot, Dumont,
  de~Bremond~d'Ars, Petrosino, Zuddas, Lopes, Boul{\'e}, Neves, {\em
  et~al.}}{Le~Mou{\"e}l {\em et~al.}}{2023}]{lemouel2023}
Le~Mou{\"e}l, J., D.~Gibert, V.~Courtillot, S.~Dumont, J.~de~Bremond~d'Ars,
  S.~Petrosino, P.~Zuddas, F.~Lopes, J.~Boul{\'e}, M.~Neves, {\em et~al.}
  (2023).
\newblock On the external forcing of global eruptive activity in the past 300
  years.
\newblock {\em Frontiers in Earth Science\/}~{\em 11}, 1254855.

\bibitem[\protect\citeauthoryear{Le~Mou{\"e}l, Gibert, Boul{\'e}, Zuddas,
  Courtillot, Lopes, G{\`e}ze, et Maineult}{Le~Mou{\"e}l {\em
  et~al.}}{2024}]{lemouel2024}
Le~Mou{\"e}l, J.-L., D.~Gibert, J.-B. Boul{\'e}, P.~Zuddas, V.~Courtillot,
  F.~Lopes, M.~G{\`e}ze, et A.~Maineult (2024).
\newblock On the effect of the luni-solar gravitational attraction on trees.
\newblock {\em arXiv preprint arXiv:2402.07766\/}.

\bibitem[\protect\citeauthoryear{Le~Mou{\"e}l, Lopes, Courtillot, et
  Gibert}{Le~Mou{\"e}l {\em et~al.}}{2019}]{lemouel2019}
Le~Mou{\"e}l, J.-L., F.~Lopes, V.~Courtillot, et D.~Gibert (2019).
\newblock On forcings of length of day changes: From 9-day to 18.6-year
  oscillations.
\newblock {\em Physics of the Earth and Planetary Interiors\/}~{\em 292},
  1--11.

\bibitem[\protect\citeauthoryear{Lemmerling et Van~Huffel}{Lemmerling et
  Van~Huffel}{2001}]{lemmerling2001}
Lemmerling, P. et S.~Van~Huffel (2001).
\newblock Analysis of the structured total least squares problem for
  hankel/toeplitz matrices.
\newblock {\em Numerical algorithms\/}~{\em 27}, 89--114.

\bibitem[\protect\citeauthoryear{Lopes, Courtillot, et Le~Mou{\"e}l}{Lopes {\em
  et~al.}}{2022}]{lopes2022}
Lopes, F., V.~Courtillot, et J.-L. Le~Mou{\"e}l (2022).
\newblock Triskeles and symmetries of mean global sea-level pressure.
\newblock {\em Atmosphere\/}~{\em 13\/}(9), 1354.

\bibitem[\protect\citeauthoryear{Maineult, Bernab{\'e}, et Ackerer}{Maineult
  {\em et~al.}}{2005}]{maineult2005}
Maineult, A., Y.~Bernab{\'e}, et P.~Ackerer (2005).
\newblock Detection of advected concentration and ph fronts from self-potential
  measurements.
\newblock {\em Journal of Geophysical Research: Solid Earth\/}~{\em
  110\/}(B11).

\bibitem[\protect\citeauthoryear{Maineult, Strobach, et Renner}{Maineult {\em
  et~al.}}{2008}]{maineult2008}
Maineult, A., E.~Strobach, et J.~Renner (2008).
\newblock Self-potential signals induced by periodic pumping tests.
\newblock {\em Journal of Geophysical Research: Solid Earth\/}~{\em 113\/}(B1).

\bibitem[\protect\citeauthoryear{Masson-Delmotte, Zhai, Pirani, Connors,
  P{\'e}an, Berger, Caud, Chen, Goldfarb, Gomis, {\em et~al.}}{Masson-Delmotte
  {\em et~al.}}{2021}]{masson2021}
Masson-Delmotte, V., P.~Zhai, A.~Pirani, S.~L. Connors, C.~P{\'e}an, S.~Berger,
  N.~Caud, Y.~Chen, L.~Goldfarb, M.~Gomis, {\em et~al.} (2021).
\newblock Climate change 2021: the physical science basis.
\newblock {\em Contribution of working group I to the sixth assessment report
  of the intergovernmental panel on climate change\/}~{\em 2\/}(1), 2391.

\bibitem[\protect\citeauthoryear{Randin, Paulsen, Vitasse, Kollas, Wohlgemuth,
  Zimmermann, et K{\"o}rner}{Randin {\em et~al.}}{2013}]{randin2013}
Randin, C.~F., J.~Paulsen, Y.~Vitasse, C.~Kollas, T.~Wohlgemuth, N.~E.
  Zimmermann, et C.~K{\"o}rner (2013).
\newblock Do the elevational limits of deciduous tree species match their
  thermal latitudinal limits?
\newblock {\em Global Ecology and Biogeography\/}~{\em 22\/}(8), 913--923.

\bibitem[\protect\citeauthoryear{Ray et Erofeeva}{Ray et
  Erofeeva}{2014}]{ray2014}
Ray, R.~D. et S.~Y. Erofeeva (2014).
\newblock Long-period tidal variations in the length of day.
\newblock {\em Journal of Geophysical Research: Solid Earth\/}~{\em 119\/}(2),
  1498--1509.

\bibitem[\protect\citeauthoryear{Saxe, Cannell, Johnsen, Ryan, et
  Vourlitis}{Saxe {\em et~al.}}{2001}]{saxe2001}
Saxe, H., M.~G. Cannell, {\O}.~Johnsen, M.~G. Ryan, et G.~Vourlitis (2001).
\newblock Tree and forest functioning in response to global warming.
\newblock {\em New phytologist\/}~{\em 149\/}(3), 369--399.

\bibitem[\protect\citeauthoryear{Svenning, Normand, et Kageyama}{Svenning {\em
  et~al.}}{2008}]{svenning2008}
Svenning, J.-C., S.~Normand, et M.~Kageyama (2008).
\newblock Glacial refugia of temperate trees in europe: insights from species
  distribution modelling.
\newblock {\em Journal of Ecology\/}~{\em 96\/}(6), 1117--1127.

\bibitem[\protect\citeauthoryear{Trumbore, Brando, et Hartmann}{Trumbore {\em
  et~al.}}{2015}]{trumbore2015}
Trumbore, S., P.~Brando, et H.~Hartmann (2015).
\newblock Forest health and global change.
\newblock {\em Science\/}~{\em 349\/}(6250), 814--818.

\bibitem[\protect\citeauthoryear{Von~Arx, Dobbertin, et Rebetez}{Von~Arx {\em
  et~al.}}{2012}]{vonarx2012}
Von~Arx, G., M.~Dobbertin, et M.~Rebetez (2012).
\newblock Spatio-temporal effects of forest canopy on understory microclimate
  in a long-term experiment in switzerland.
\newblock {\em Agricultural and Forest Meteorology\/}~{\em 166}, 144--155.

\bibitem[\protect\citeauthoryear{Zellweger, De~Frenne, Lenoir, Vangansbeke,
  Verheyen, Bernhardt-R{\"o}mermann, Baeten, H{\'e}dl, Berki, Brunet, {\em
  et~al.}}{Zellweger {\em et~al.}}{2020}]{zellweger2020}
Zellweger, F., P.~De~Frenne, J.~Lenoir, P.~Vangansbeke, K.~Verheyen,
  M.~Bernhardt-R{\"o}mermann, L.~Baeten, R.~H{\'e}dl, I.~Berki, J.~Brunet, {\em
  et~al.} (2020).
\newblock Forest microclimate dynamics drive plant responses to warming.
\newblock {\em Science\/}~{\em 368\/}(6492), 772--775.

\end{thebibliography}

\end{document}